\documentclass[aip,twocolumn,
reprint,amsmath,amssymb]{revtex4-1}
\usepackage{graphicx}% Include figure files
\usepackage{dcolumn}% Align table columns on decimal point
\usepackage{bm}% bold math
\usepackage{float}
\usepackage{subfigure}
\usepackage{color,soul}
\usepackage{gensymb}
\usepackage{lipsum}
\usepackage{amsmath}
\usepackage{amsfonts}
\usepackage{lineno}
\usepackage{cancel}
\usepackage[normalem]{ulem}

\usepackage[section]{placeins}   %Avoid table or figure floating body crossing the section

\begin{document}
\preprint{}

%\linenumbers

\title{Generation of polarized electron beams through self-injection in the interaction of a laser with a pre-polarized plasma}

\author{L. R. Yin}
\affiliation{Key Laboratory of Nuclear Physics and Ion-beam Application (MOE), Institute of Modern Physics, Department of Nuclear Science and Technology, Fudan University, Shanghai 200433, China}
\author{X. F. Li} \email{xiaofengli@siom.ac.cn}
\affiliation{State Key Laboratory of High Field Laser Physics, Shanghai Institute of Optics and Fine Mechanics, Chinese Academy of Sciences, Shanghai 201800, China.}
\author{Y. J. Gu}
\affiliation{SANKEN, Osaka University, Mihogaoka 8-1, Ibaraki, Osaka 567-0047, Japan}
\author{N. Cao}
\affiliation{Sichuan Research Institute, Shanghai Jiao Tong University, Sichuan 610200, China}
\author{Q. Kong}\email{qkong@fudan.edu.cn}
\affiliation{Key Laboratory of Nuclear Physics and Ion-beam Application (MOE), Institute of Modern Physics, Department of Nuclear Science and Technology, Fudan University, Shanghai 200433, China}
\author{M. B\"uscher}
\affiliation{Peter Gr\"unberg Institut (PGI-6), Forschungszentrum J\"ulich, Wilhelm-Johnen-Str. 1, 52425 J\"ulich, Germany}
\affiliation{Institut f\"ur Laser- und Plasmaphysik, Heinrich-Heine-Universit\"at D\"usseldorf, Universit\"atsstr. 1, 40225 D\"usseldorf, Germany}
\author{S. M. Weng}
\affiliation{Key Laboratory for Laser Plasmas (MoE), School of Physics and Astronomy, Shanghai Jiao Tong University, Shanghai 200240, China}%
\affiliation{Collaborative Innovation Center of IFSA, Shanghai Jiao Tong University, Shanghai 200240, China}
\author{M. Chen}
\affiliation{Key Laboratory for Laser Plasmas (MoE), School of Physics and Astronomy, Shanghai Jiao Tong University, Shanghai 200240, China}%
\affiliation{Collaborative Innovation Center of IFSA, Shanghai Jiao Tong University, Shanghai 200240, China}
\author{Z. M. Sheng}
\affiliation{Key Laboratory for Laser Plasmas (MoE), School of Physics and Astronomy, Shanghai Jiao Tong University, Shanghai 200240, China}%
\affiliation{Collaborative Innovation Center of IFSA, Shanghai Jiao Tong University, Shanghai 200240, China}%
\affiliation{Tsung-Dao Lee Institute, Shanghai Jiao Tong University, Shanghai 200240, China}

\date{\today}%15 July 2020

\begin{abstract}

Polarized electron beam production via laser wakefield acceleration in pre-polarized plasma is investigated by particle-in-cell simulations. The evolution of the electron beam polarization is studied based on the Thomas-Bargmann-Michel-Telegdi equation for the transverse and longitudinal self-injection, and the depolarization process is found to be influenced by the injection schemes. In the case of transverse self-injection as found typically in the bubble regime, the spin precession of the accelerated electrons is mainly influenced by the wakefield. However, in the case of longitudinal injection in the quasi-one-dimensional regime (for example, F. Y. Li \emph{et al}., Phys. Rev. Lett. 110, 135002 (2013)), the direction of electron spin oscillates in the laser filed. Since the electrons move around the laser axis, the net influence of the laser field is nearly zero and the contribution of the wakefield can be ignored. Finally, an ultra-short electron beam with polarization of $99\%$ can be obtained using longitudinal self-injection.

\end{abstract}

\maketitle

\section{INTRODUCTION}
 As an advanced accelerator method, laser wakefield accelerators (LWFA)\cite{T. Tajima1979,E. Esarey2009} have been developing steadily both theoretically and experimentally in the recent decades\cite{A. Gonsalves2019,C. Geddes2004,J. Faure2004,S. P. Mangles2004,X. Wang2013,M. H. Cho2018,J. Osterhoff2008, A. Buck2013,E. Brunetti2010,R. Weingartner2012,W. P. Leemans2006} owing to the rapidly advancing laser technology, especially chirped-pulse amplification\cite{G. Mourou1992}. A variety of mechanisms has been proposed to control the electron beams properties which are comparable with those from conventional particle accelerators, such as energy spectra\cite{A. Gonsalves2019,S. P. Mangles2004,J. Faure2004,X. Wang2013,C. Geddes2004}, controllability\cite{M. H. Cho2018}, stability\cite{J. Osterhoff2008, A. Buck2013}, beam emittance\cite{E. Brunetti2010,R. Weingartner2012} and beam energy\cite{W. P. Leemans2006}. Many efforts have been made with controllable injection mechanisms to improve the electron beam quality, such as density-transition injection\cite{S. Bulanov1998,J. Wang2022}, ionization-induced injection\cite{A. Pak2010}, or colliding-pulse injection\cite{E. Esarey1997}. However, the control of electron beam polarization has not been investigated thoroughly.

Spin-polarized electron beams have been widely used in material science\cite{S.-Y. Lee2018,P. J. Schultz1988}, particle and nuclear physics\cite{C. Glashausser1979,J. R. Danielson2015,G. Moortgat-Pick2008,V. Shiltsev2021}. Such beams are generally produced by the radiative polarization due to the Sokolov-Ternov effect in conventional accelerators\cite{A. A. Sokolov1967,S. R. Mane2005} i.e., storage rings, which takes about a few hours in polarization build-up. In contrast, the acceleration process can be accomplished within a few picoseconds in a plasma accelerator. In $2017$, a pre-polarized gas plasma has been produced through laser-induced photo-dissociation in experiment\cite{D. Sofikitis2017}. Later, Wen \emph{et al.} proposed to generate high-current polarized electron beams with $90.6\%$ spin polarization through LWFA based on the density-transition injection mechanism\cite{M. Wen2019}. The study of Nie \emph{et al.} shown that an electron beam with up to $\sim56\%$ polarization could be obtained using the ionization-induced injection mechanism\cite{Z. Nie2021,Z. Nie2022}. More recently, the effect of bubble geometry on polarization of self-injection electrons has been studied. It was found that the deviation from a perfect spherical symmetry severely degrades the polarization of electron beam during the transverse injection\cite{H. C. Fan2022}. Recently, Gong \emph{et al.} proposed that the colliding-pulse injection scheme enables the production of quasi-monoenergetic electron beams in excess of $80\%$ polarization and tens of pC charge with commercial $10$TW laser systems in a pre-polarized plasma\cite{Z. Gong2023,S. Bohlen2023}. Furthermore, Sun \emph{et al.} proposed to generate an attosecond electron bunches with polarization $\sim90\%$ through using a radially polarized laser interacted with a pre-polarized plasma\cite{T. Sun2022}. Moreover, an energetic spin-polarization electron beams can also be produced by vortex Laguerre-Gaussian laser\cite{Y. T. Wu2019} or beam-driven wakefield acceleration\cite{Y. T. Wu2 2019}.

While these injection mechanisms have been investigated to control the polarization on LWFA electron beam\cite{M. Wen2019,Z. Nie2021,Z. Nie2022,H. C. Fan2022,Z. Gong2023,S. Bohlen2023,T. Sun2022}, the self-injection mechanism with relative simple setup still needs to be analyzed thoughtfully.
There are two self-injection schemes, transverse and longitudinal, as demonstrated in Ref. [37]. %\cite{S. Corde2013}.
The transverse injection mainly happens in the $3$D nonlinear bubble regime. The accelerated electrons initially stay away from the laser axis, move in the bubble sheath, arrive at the tail of bubble and are injected in wakefield\cite{S. V. Bulanov1997,A. Pukhov2002,S. Corde2013}. However, the trajectories of the accelerated electrons in the longitudinal injection scheme are different, which mainly takes place in the quasi-1D regime of wakefield\cite{F. Y. Li2013}. The electrons initially located at the front of the laser pulse slip backward along the laser axis after interacting with the laser. Once reaching the tail of wakefield, the electrons are injected and finally accelerated by the wakefield\cite{F. Y. Li2013,M. K. Weikum2016,S. Corde2013}.

Previous studies\cite{M. Wen2019,Z. Nie2021,Z. Nie2022,H. C. Fan2022,Z. Gong2023,S. Bohlen2023,T. Sun2022} have shown that the properties of the electron beams depend on the electron injection mechanism. The electron polarization mainly changes during the injection process. In this paper, we study the polarization of the electron beam for the longitudinal injection scheme in a fully pre-polarized plasma with an up-ramp-plateau density profile. The longitudinal scheme is found to be more beneficial in generating high spin polarization electron beams as compared to the transverse case. Our work is divided into three sections. Section II introduces the simulation setup with a brief description about the longitudinal injection scheme. In Sec. III, we present numerical results and a discussion. The conclusions can be found in Sec. IV.

\section{SIMULATION Method}

\begin{figure}[t]
	\centering
	\includegraphics[width=0.39\textwidth]{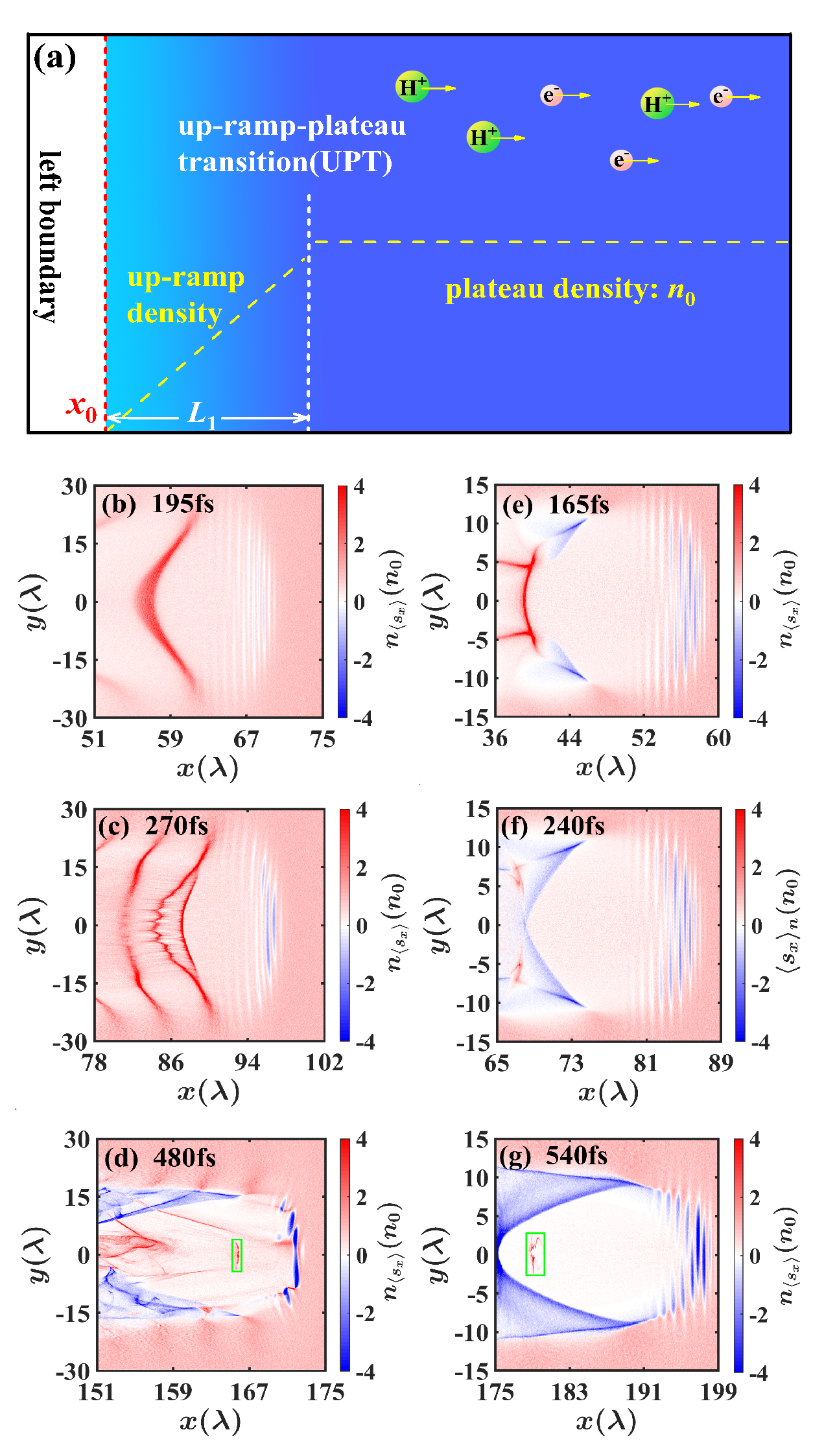}
	\caption{(a) Schematic representation of the initially pre-polarized plasma. The longitudinal profile of the electron density is marked by the yellow dashed line including an up-ramp from $0$ to $n_{0}$ with length $L_{1}$ and a plateau with $n_{0}$. The initial polarization direction is aligned along the $x$ direction as denoted by the arrows. The laser is focused at the left boundary of the plasma ($x_{0}=30\lambda$). For the case of longitudinal injection (Case A), (b)-(d) show the density distribution of electron longitudinal polarization ($n_{\langle s_{x}\rangle}$) at three different times, i.e., the product of electron density (normalized by $n_{0}$) and their average of polarization in the $x$ direction ($\langle s_{x}\rangle$) per cell. Here, $a_{0}=6$, $\tau=17\mathrm{fs}$, $w_{0}=20\lambda$, $n_{0}=0.04n_{c}$ and $L_{1}=45\lambda$. For the case of transverse injection (Case B), (e)-(f) present the corresponding distributions of $n_{\langle s_{x}\rangle}$ at different times, where $w_{0}=10\lambda$, $n_{0}=0.01n_{c}$ and $L_{1}=10\lambda$. The other parameters are the same as Case A. The electrons with kinetic energy $E_{k}>13\mathrm{MeV}$ are chosen as the acceleration electrons, which are marked by a green box in (d) and (g), respectively.} \label{fig1}	
\end{figure}

In this study, two-dimensional ($2$D) PIC simulations were performed with a modified version of the EPOCH code\cite{T. D. Arber2015}, which includes the spin evolution module based on the Thomas-Bargmann-Michel-Telegdi(TBMT)\cite{L. H. Thomas1926,S. Mane2005} equation via the Boris pusher method \cite{X. F. Li2021}. The electron spin is regarded as a quasiclassical quantity with a vector $\boldsymbol{s}$, which has an absolute value of $1$ and a direction calculated from the TBMT equation
$\mathrm{d}\boldsymbol{s}/\mathrm{d}t=\boldsymbol{\Omega}\times \boldsymbol{s}$ with
\begin{equation}\label{eq2}
\begin{split}
\boldsymbol{\Omega}=\frac{e}{m_{e}}\Bigg[&\left(\boldsymbol{a_{e}}+\frac{1}{\gamma}\right)\boldsymbol{B}-\frac{\boldsymbol{a_{e}}\gamma}{\gamma+1}\boldsymbol{v}\cdot \boldsymbol{B}\frac{\boldsymbol{v}}{c^{2}}
\\&-\left(\boldsymbol{a_{e}}+\frac{1}{\gamma+1}\right)\frac{\boldsymbol{v}}{c^{2}}\times\boldsymbol{E}\Bigg],
\end{split}
\end{equation}
where $m_{e}$, $e$ and $\boldsymbol{a_{e}}\approx1.16\times10^{-3}$ are the electron mass, charge and the dimensionless anomalous magnetic moment, respectively, $\gamma$ is the Lorentz factor of the electron, $c$ is the light speed in vacuum, $\boldsymbol{B}$ is the magnetic field, and $\boldsymbol{E}$ is the electric field in the laboratory frame. The effects of radiation reaction, Stern-Gerlach and Sokolov-Ternov can be ignored during the study of LWFA, based on the work of Thomas \emph{et al.} \cite{J. Thomas2020}.

In the simulation, the laser propagates in the $x$-direction with linear polarization and a Gaussian envelope in the $y$-direction
\begin{equation}\label{eq1}
E=\frac{E_{0}w_{0}}{w\left(x\right)}\mathrm{exp}\left[-\frac{y^{2}+z^{2}}{w\left(x\right)^{2}}-\frac{\left(kx-\omega t\right)^{2}}{\left(0.5\tau\right)^{2}}\right]\!\cos\left(\varphi\right)
\end{equation}
with the laser wavelength $\lambda=800$$\mathrm{nm}$, the initial laser waist $w_{0}=20\lambda$, $w(x)=w_{0}\left[1+\left(x-x_{0}\right)^{2}/z_{R}^{2}\right]^{0.5}$, $z_{R}=\pi w_{0}^{2}/\lambda$, the pulse duration $\tau=17 \mathrm{fs}$ and the normalized laser amplitude $a_{0}=eE_{0}/m_{e}\omega c=6$, corresponding to a peak intensity of $I_{0}=7.71\times10^{19}\mathrm{W/cm}^{2}$. The simulation box is $200\lambda(x)\times120\lambda(y)$ with resolution $\mathrm{d}x=0.02\lambda$ and $\mathrm{d}y=0.08\lambda$. Open boundary conditions are used in both directions and there are $4$ pseudo-particles per cell for each particle species.

The initial longitudinal profile of the pre-polarized plasma is an up-ramp followed by a plateau with constant density $n_{0}=0.04n_{c}$, as shown in Fig. \ref{fig1}(a), marked as yellow dashed line. Such a density profile enable the longitudinal electron injection possible, as first introduced in Ref. [40].
Here, the length of the up-ramp transition is $L_{1}=45\lambda$ and the laser pulse is focused on the left edge of the plasma target at $x_{0}=30\lambda$. The pre-polarized plasma could be realized by using ultra-violet polarization method\cite{D. Sofikitis2017}. For simplicity, the initial polarization rate of the plasma is assumed as $100\%$, where we are interested in the evolution of the polarization during self-injection scheme. The net polarization of a particle beam is defined as $P=\sqrt{\left\langle s_{x} \right\rangle^{2}+\left\langle s_{y} \right\rangle^{2}+\left\langle s_{z} \right\rangle^{2}}$, where $s_{i}$ is the components of spin polarization in each direction and $\left\langle s_{i} \right\rangle$ is the corresponding average value.

\section{results \& discussions}
Generally, the profile of the wakefield depends on the parameters of the laser and plasma, which inevitably causes the variation of self-injection scheme, futher leading to different evolution of spin polarization during self-injection process. When the laser spot size is larger than the plasma wavelength (Case A), the wakefield is a quasi-$1$D regime, as shown as in Fig. \ref{fig1}(b). At this time, the wakefield propagates in the up-rump density. When the wakefield reaches in the uniform density regime ($x=75\lambda$), several electrons located at the tail of wakefield ($x=88\lambda$), can be captured and accelerated due to breaking-wave effect, as presented in Fig. \ref{fig1}(c). After that, owning to the effect of laser self-focusing, the laser intensity increases, the wakefield develops into a $3$D nonlinear bubble regime and the electrons are accelerated continually, as revealed as Fig. \ref{fig1}(d). On the other hand, when the laser spot size is equal to the plasma wavelength and the laser intensity $a_{0}$ is larger than $4$, \emph{i.e.}, Case B, the bubble regime can be formed directly as the laser propagated into the plasma. In order to avoid the electron at the left boundary injected into the bubble, an up-rump density with a short length is also used, as shown as in Fig. \ref{fig1}(f). Different to the case of longitudinal injection, the bubble regime is a $3$D nonlinear regime initially and the corresponding phase velocity slows down. As the bubble geometry changes following the laser evolution, several electrons can be injected in the bubble and achieve acceleration, as shown as in Fig. \ref{fig1}(g).

\begin{figure}[t]
\centering
\includegraphics[width=0.48\textwidth]{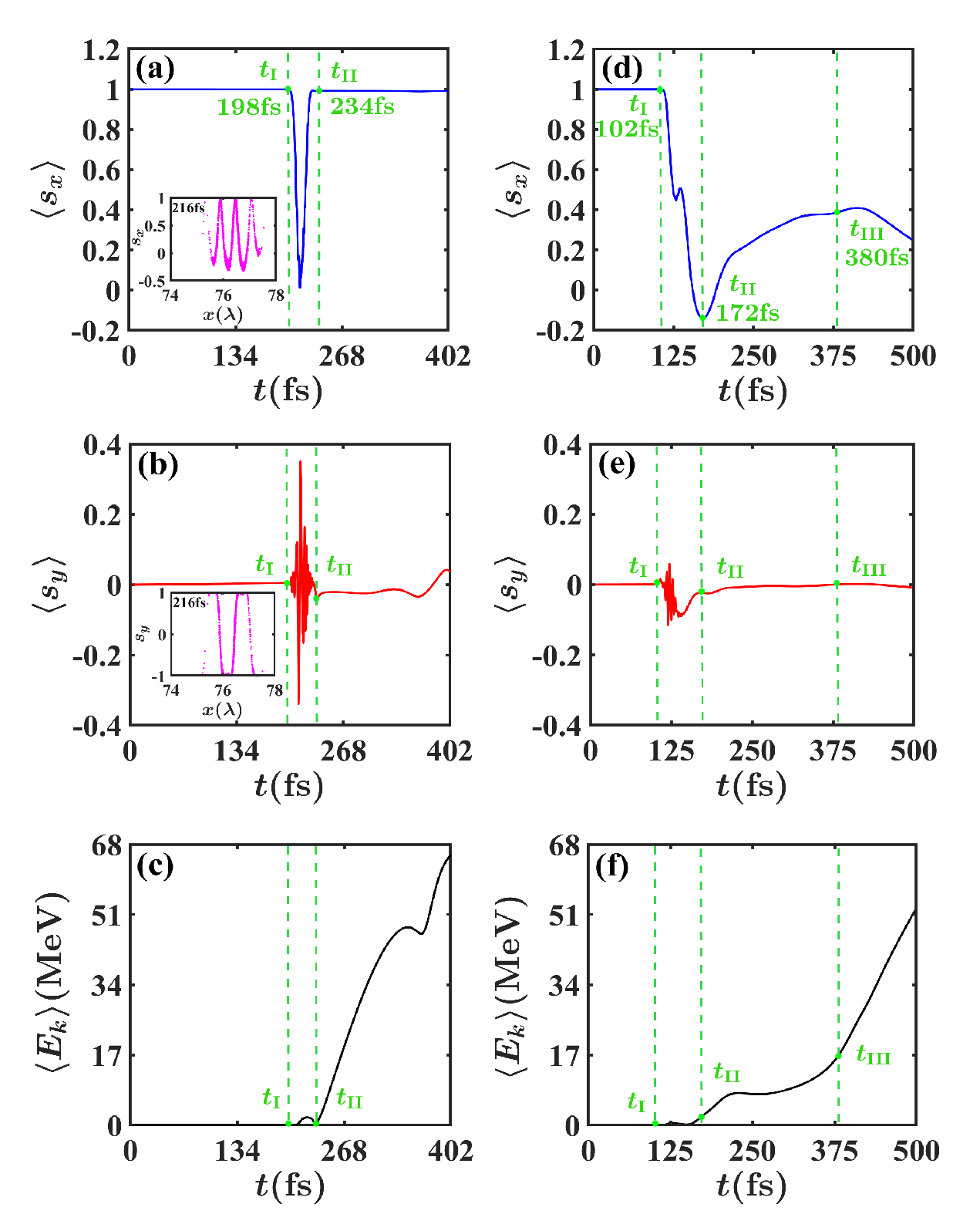}
\caption{The history of particle properties, $\left\langle s_{x}\right\rangle$ (a), $\left\langle s_{y}\right\rangle$ (b) and the average kinetic energy $\left\langle E_{k}\right\rangle$ (c), about acceleration electrons in the case of the longitudinal scheme. The distribution of $s_x$ (or $s_y$) in the x-direction is shown in the insert of (a) (or (b)). (d)-(f) show the corresponding quantities in the case of transverse scheme. The accelerated electrons are marked in the Figs. \ref{fig1}(d) and \ref{fig1}(g), respectively.}\label{fig2}
\end{figure}

Not surprisingly, the distribution of the electron polarization is different in the two cases and the influence of laser on the electron polarization cannot be ignored. In the quasi-$1$D regime (Case A), the values of $s_{x}$ for injection electrons mostly are positive, as plotted in Fig. \ref{fig1}(c). While in the bubble regime (Case B), the value of $s_{x}$ about the electrons located at the sheath are negative, as revealed in Fig. \ref{fig1}(f). For further analysis, the accelerated electrons are chosen to analyze the evolution of polarization, as marked in the green region in Figs. \ref{fig1}(d) and \ref{fig1}(g), respectively. For Case A, $6056$ electrons are chosen with a polarization $P=0.99$. For Case B, there are $32078$ electrons chosen with a polarization $P=0.18$.

\begin{figure}[t]
\centering
\includegraphics[width=0.46\textwidth]{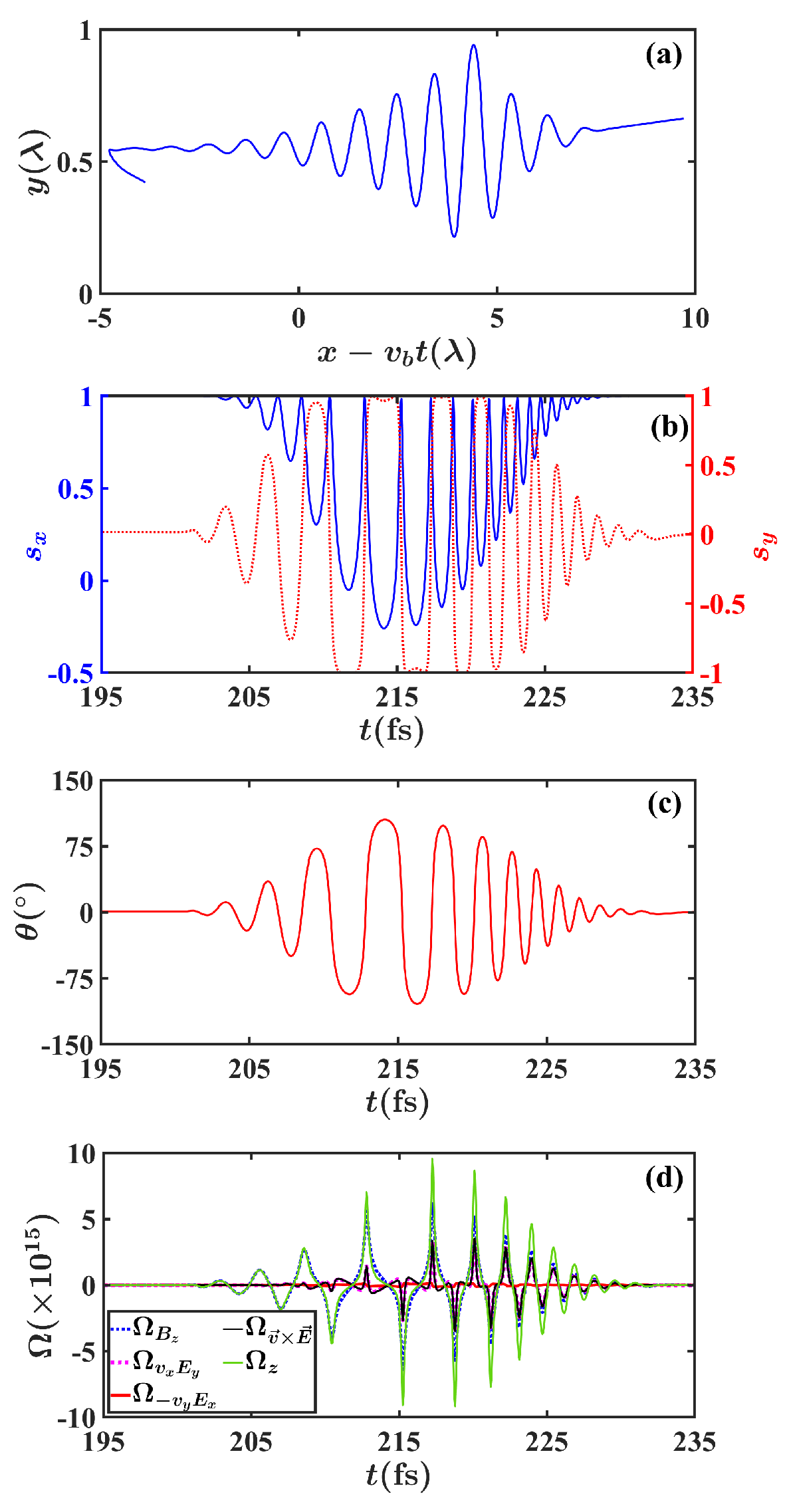}
\caption{(Color online) (a) Trajectory of a typical tracked electron for the longitudinal self-injection scheme (Case A) at wakefield frame, where $v_{b}$ is the phase velocity of wakefield calculated using plateau density. The electron is located at the front of wakefield. (b) The history of $s_{x}$ (blue solid line) and $s_{y}$ (red dashed line) for the tracked electron. (c) The evolution of spin direction ($\theta=\tan^{-1}(s_{y}/s_{x})$) with time. (d) The evolution of $\boldsymbol{\Omega}_{z}$ (green solid line), term $\boldsymbol{\Omega}_{B_{z}}$ (blue dashed line), term $\boldsymbol{\Omega}_{v_{x}E_{y}}$ (magenta dashed line) of $\boldsymbol{\Omega}_{\boldsymbol{v}\times\boldsymbol{E}}$ (black solid line) caused by $v_{x}E_{y}$ and term $\boldsymbol{\Omega}_{-v_{y}E_{x}}$ (red solid line) of $\boldsymbol{\Omega}_{\boldsymbol{v}\times\boldsymbol{E}}$ caused by $-v_{y}E_{x}$ for the tracked electron.}\label{fig3}
\end{figure}

The history of $\left\langle s_{x} \right\rangle$, $\left\langle s_{y} \right\rangle$ and the average energy $\left\langle E_{k} \right\rangle$ for these two cases are plotted in Fig. \ref{fig2}. As presented in Figs. \ref{fig2}(a)-\ref{fig2}(c), the evolution of the polarization for the longitudinal injection can be divided into three stages: (i) $t<t_I=198 \mathrm{fs}$, the electron fixed. (ii) $t_I<t<t_{II}=234 \mathrm{fs}$, the value of $\left\langle s_{x} \right\rangle$ decreases firstly and then returns nearly initial value during a short time, which is nearly the duration of laser ($\sim36 \mathrm{fs}$). As presented in the insert of Fig. \ref{fig2}(a), the distribution of $s_x$ shown that the laser can affect the electron spin directly at $216 \mathrm{fs}$. The oscillation period is nearly $0.5\lambda$ for $\left\langle s_{x} \right\rangle$ and $1\lambda$ for $\left\langle s_{y} \right\rangle$, which means that the electron spin is affected by the laser field directly. Meanwhile, the average energy increases firstly and decreases. (iii) $t>t_{II}$, the value of $\left\langle s_{x} \right\rangle$ does not change obviously. The value of $\left\langle s_{y} \right\rangle$ is nearly $0$ owning to the azimuthal symmetry of the wakefield. At this stage, the electron energy increases following with time and it means they are continuously accelerated in the wakefield.

For the case of the transverse injection scheme (Case B), as analyzed in Ref. [31]%\cite{H. C. Fan2022}
, the evolution of polarization can be divided into four stages: (i) $t<t_{I}$, the electrons do not feel the wakefield. (ii) $t_{I}<t<t_{II}$, the electrons are located on the bubble shell. $\left\langle s_{x} \right\rangle$ decreases and $\left\langle s_{y} \right\rangle$ oscillates in the laser field and stays nearly $0$ due to the azimuthal symmetry of the bubble field, as shown in Fig. \ref{fig2}(d). (iii) $t_{II}<t<t_{III}$, the electrons reach the tail of the bubble and $\left\langle s_{x} \right\rangle$ increases as revealed in Fig. \ref{fig2}(d). (iv) $t>t_{III}$, the electrons are captured in the bubble and their spin precession slows down.

The electron spin evolution during the transverse injection has been studied through single particle dynamics in the work of Fan \emph{et al.}\cite{H. C. Fan2022} It is found that the electron spin is mainly affected by the magnetic field of the bubble during the second stage and affected by the electric field of the bubble at the third stage. At the fourth stage, the electron moves along with the laser axis, so its spin does not change obviously. For the longitudinal injection, a typical accelerated electron is also analyzed, as shown as in Fig. \ref{fig3}. The trajectory of the electron in the wakefield coordinate system between $195 \mathrm{fs}$ and $235 \mathrm{fs}$ is presented in Fig. \ref{fig3}(a). At $195 \mathrm{fs}$, the electron locates at the head of wakefield, then it slips backward. When reaching at the tail of the wakefield, it is captured. Although it vibrates at the transverse direction, the transverse position does not change obviously in the wakefield.

Moreover, the evolutions of $s_{x}$ and $s_{y}$ are plotted in Fig. \ref{fig3}(b). Based on the TBMT equation, the electron spin processes in the XY plane with the laser field. Its spin changes rapidly in the laser field and the $s_{x}$ returns to its initial value at each cycle. The amplitude of oscillation is coincided with laser intensity and the profile of $s_{x}$ is similar with the laser duration. However, the period of oscillation increases firstly and then decreases. If we defined the spin angle $\theta=\tan^{-1}(s_{y}/s_{x})$, the quiver of electron spin is presented more clearly in Fig. \ref{fig3}(c). The electron spin oscillates around the x-direction in the laser field and it caused that the oscillation period of $s_{y}$ is twice as that of $s_{x}$. Additionally, the period of oscillation decreases with time.

\begin{figure}[t]
	\centering
	\includegraphics[width=0.48\textwidth]{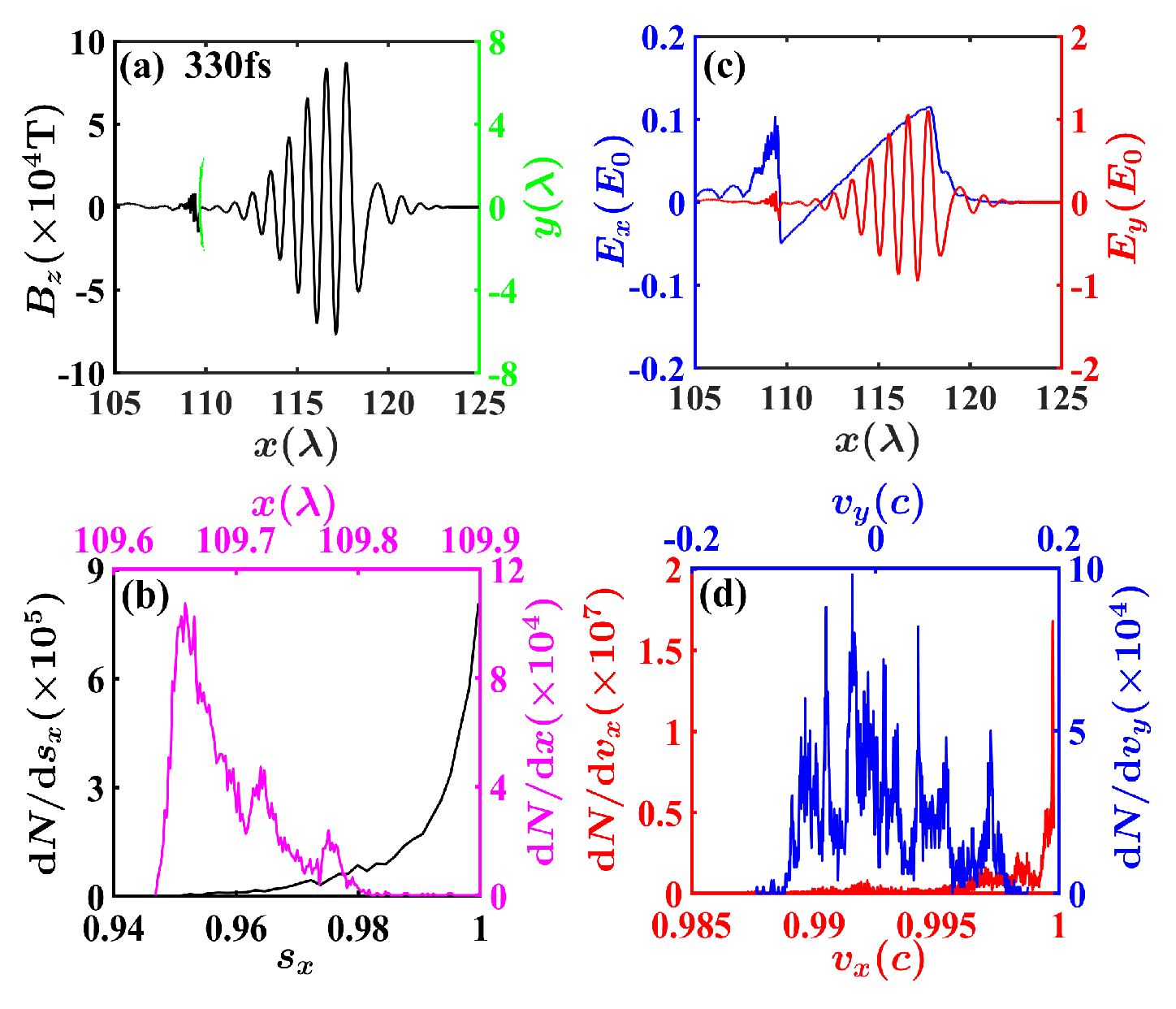}
	\caption{(a) The green dots denote the positions of the chosen accelerated electrons, which have been marked in Fig. \ref{fig1}(d). The magnetic field $B_{z}$ at the laser axis is presented as black solid line. (b) The spectra of $s_{x}$ (black line) and the longitudinal position $x$ (magenta line) for the accelerated electrons at $330 \mathrm{fs}$. (c) The profiles of $E_{y}$ (red line) and $E_{x}$ (blue line) at the laser axis. (d) The spectra of the longitudinal velocity $v_{x}$ (blue line), transverse velocity $v_{y}$ (red line).}\label{fig4}
\end{figure}

In order to investigate the dynamics of the electron spin, the contribution of the electromagnetic field on the precession frequency $\boldsymbol{\Omega}$ is analyzed in detail. Considering $\boldsymbol{a_{e}}\approx1.16\times10^{-3}$ for the electron, the contribution term of $\boldsymbol{\Omega}$ can be simplified as
 \begin{equation}
	\boldsymbol{\Omega}\simeq \frac{e}{m_{e}}\left(\frac{1}{\gamma}\boldsymbol{B}
	-\frac{1}{\gamma+1}\frac{\boldsymbol{v}}{c^{2}}\times\boldsymbol{E}\right).
 \end{equation}
As shown in Fig. \ref{fig3}(c), the electron precesses in the XY plane, which is caused by the part of $\boldsymbol{\Omega}_{z}$. It can be divided as three terms, $\boldsymbol{\Omega}_{B_{z}}$, $\boldsymbol{\Omega}_{-v_{y}E_{x}}$ and $\boldsymbol{\Omega}_{v_{x}E_{y}}$. Figure \ref{fig3}(d) presents the history of $\boldsymbol{\Omega}_{z}$ and the contribution of three different terms have been compared. It is found that the contribution of $\boldsymbol{\Omega}_{-v_{y}E_{x}}$ can be ignored because the motion of electron is near along the laser axis, where $E_{x}$ is nearly zero. The contribution of $\boldsymbol{\Omega}_{v_{x}E_{y}}$ is comparable with $\boldsymbol{\Omega}_{\boldsymbol{v}\times\boldsymbol{E}}$. Following with time, the contribution of $\boldsymbol{\Omega}_{\boldsymbol{v}\times\boldsymbol{E}}$ increases firstly and then decreases, which is consistent with the evolution of electron velocity. More importantly, the net contribution of $\boldsymbol{\Omega}_{z}$ is nearly zero at one cycle, such that the effect of laser field on electron spin can be ignored.

As revealed in Fig. \ref{fig2}(a), in the third stage, the electron spin do not change obviously. The distribution of the magnetic field at $330$fs is shown in Fig. \ref{fig4}(a). The accelerated electrons are denoted as green dots and they are located at the tail of the wakefield. The spectrum of $s_{x}$ indicates that the spin does not change obviously compared with the initial value as presented in Fig. \ref{fig4}(b), which means that the effect of the wakefield can be ignored. Furthermore, the width of the accelerated electron beam is $0.039\lambda$ (or $103.20$as), where the full width at half maximum (FWHM) of the energy spectrum was used. The distribution of the electromagnetic field at the laser axis is presented in Fig. \ref{fig4}(c) and the spectra of $v_{x}$ and $v_{y}$ are plotted in Fig. \ref{fig4}(d). At the tail of the wakefield, $B_{z}$ and $E_{y}$ are nearly zero. Then their contribution to $\boldsymbol{\Omega}_{z}$ can be ignored. Because of the small $v_{y}$, the contribution of $\boldsymbol{\Omega}_{-v_{y}E_{x}}$ can also be ignored, which is an essential difference to the transverse injection. In the case of longitudinal injection, due to the motion of electrons close to the laser axis, the electron spin is affected by the laser field only. Since the laser field cannot depolarize the electrons, it is advantageous to obtain a polarized electron beam with a high energy in LWFA. Finally, a nearly $115 \mathrm{as}$ electron beam with $99\%$ polarization and average kinetic energy $64 \mathrm{MeV}$ is obtained at $400 \mathrm{fs}$.

\section{summary}
We have studied the generation of an electron beams including its polarization properties in the bubble regime of LWFA. By using a series of $2$D PIC simulations, it is found that the depolarization process depends on the self-injection scheme. Compared with transverse self-injection, the longitudinal self-injection is more suitable to generate an electron beam with higher polarization. The accelerated electrons move around the laser axis in the case of longitudinal injection. It causes that the motion and the spin of the electrons oscillate in the laser field and the net influence of the laser field can be ignored. Moverover, the contribution of the bubble field on the spin precession is also negligible, since the transverse electromagnetic field and the transverse velocity of the electrons are both very small. Ultimately, an attosecond electron beam with polarization of $99\%$ is obtained in the simulation. Our work helps to generate a polarized electron beam using the longitudinal self-injection scheme in a pre-polarized plasma and guide the future experiments for producing the ultra-short electron beams with high polarization.

\section{ACKNOWLEDGEMENTS}
 This work was supported by the National Natural Science Foundation of China (No.11804348, No.11775056, No.11975154, and No.11991074), and the Science Challenge Project (No.TZ2018005). The work of M.B. was carried out in the framework of the J\"ulich Short-Pulse Particle and Radiation Center \cite{buscher2020jusparc} and was supported by the Accelerator Technology Helmholtz Infrastructure consortium ATHENA.

\section{DECLARATION OF INTEREST}
The authors report no conflict of interest.\\

\end{document}